\documentclass[conference]{IEEEtran}
\IEEEoverridecommandlockouts
\usepackage{cite}
\usepackage{amsmath,amssymb,amsfonts}
\usepackage{algorithmic}
\usepackage{graphicx}
\usepackage{textcomp}
\usepackage[dvipsnames]{color}
\usepackage{cite}
\usepackage{amsmath,amssymb,amsfonts}
\usepackage{algorithmic}
\usepackage{algorithm}
\usepackage{graphicx}
\usepackage{textcomp}
\usepackage{bm}
\usepackage{amsmath,amsfonts,amssymb,amsbsy,nccmath} \usepackage{amsthm}

\usepackage[utf8]{inputenc}

\usepackage{amsmath}
\usepackage{accents}
\newlength{\dhatheight}

\newcommand{\bma}{\mathbf a}

\newcommand{\bmP}{\mathbf P}

\newcommand{\bmH}{\mathbf H}

\newcommand{\bmV}{\mathbf V}

\newcommand{\bmY}{\mathbf Y}

\newcommand{\bmq}{\mathbf q}

\newcommand{\bmp}{\mathbf p}

\usepackage{xspace}
\usepackage{amssymb}
\usepackage{graphicx}
\usepackage{cite}
\usepackage{dblfloatfix}  
\usepackage{bbm}
\usepackage{dblfloatfix} 
\usepackage{xcolor}
\usepackage{setspace}
\usepackage[normalem]{ulem}
\usepackage[user]{zref}
\newcommand{\sot}[1]{} 

\usepackage{caption}
\usepackage{mathptmx}
\usepackage[scaled=.90]{helvet}

\DeclareCaptionFont{xipt}{\fontsize{8}{13}\mdseries}
\usepackage[font=xipt]{caption}

\newcounter{revc}
\makeatletter \zref@newprop{revcontent}{} \zref@addprop{main}{revcontent}
\zref@newprop{revsec}{} \zref@addprop{main}{revsec}
\zref@newprop{revpage}{} \zref@addprop{main}{revpage}

\newcommand{\revi}[2]{
\zref@setcurrent{revsec}{\thesection}%
\zref@setcurrent{revpage}{\thepage}%
\zref@setcurrent{revcontent}{#2}%
\refstepcounter{revc}%
\label{#1}%
\zlabel{#1}%
\textcolor{blue}{#2}%
}

\newcommand{\revinu}[2]{%
\zref@setcurrent{revsec}{\thesection}%
\zref@setcurrent{revcontent}{#2}%
\refstepcounter{revc}%
\zlabel{#1}%
\label{#1}
#2 }

\newcommand{\revr}[2]{%
\zref@setcurrent{revsec}{\thesection}%
\zref@setcurrent{revcontent}{#2}%
\refstepcounter{revc}%
\zlabel{#1}%
\label{#1} \sot{#2}} \makeatother

\expandafter\def\expandafter\quote\expandafter{\quote\onehalfspacing\fontsize{12}{14}\selectfont}

\usepackage[framemethod=tikz]{mdframed}
\usepackage{lipsum}

\definecolor{mycolor}{rgb}{0.122, 0.435, 0.698}

\newmdenv[innerlinewidth=0.5pt, roundcorner=4pt,linecolor=mycolor,innerleftmargin=6pt,
innerrightmargin=6pt,innertopmargin=6pt,innerbottommargin=6pt]{mybox}

\begin{document}
\title{Signal Processing Based Antenna Pattern Characterization for MIMO Systems}

\author{Chandan~Kumar~Sheemar,
   \;Jorge Querol,  
and Symeon Chatzinotas\\ 
\IEEEauthorblockA{
Interdisciplinary Centre for Security, Reliability and Trust (SnT), University of Luxembourg, Luxembourg \\email: \{chandankumar.sheemar jorge.querol  symeon.chatzinotas\}@uni.lu
}
}
 
\maketitle

\begin{abstract}
Sophisticated antenna technologies are constantly evolving to meet the escalating data demands projected for 6G and future networks. The characterization of these emerging antenna systems poses challenges that necessitate a reevaluation of conventional techniques, which rely solely on simple measurements conducted in advanced anechoic chambers. In this study, our objective is to introduce a novel endeavour for antenna pattern characterization (APC) in next-generation multiple-input-multiple-output (MIMO) systems by utilizing the potential of signal processing tools. In contrast to traditional methods that struggle with multi-path scenarios and require specialized equipment for measurements, we endeavour to estimate the antenna pattern by exploiting information from both line-of-sight (LoS) and non-LoS contributions. This approach enables antenna pattern characterization in complex environments without the need for anechoic chambers, resulting in substantial cost savings. Furthermore, it grants a much wider research community the ability to independently perform APC for emerging complex 6G antenna systems, without relying on anechoic chambers. Simulation results demonstrate the efficacy of the proposed novel approach in accurately estimating the true antenna pattern.
\end{abstract}

 
\IEEEpeerreviewmaketitle

\section{Introduction}
Emerging wireless networks are revolutionizing the way we connect and communicate in the digital age. With the rapid advancement of technology, these networks are pushing the boundaries of connectivity, speed, and reliability. From 5G networks that offer lightning fast data transfer speeds and low latency \cite{sheemar2021game,thomas2019multi} to the exciting potential of 6G networks on the horizon, emerging wireless networks promise to deliver seamless connectivity to a growing range of devices, including smart homes, autonomous vehicles, and the Internet of Things (IoT) \cite{jiang2021road,sheemar2020receiver}. These networks also aim to enhance the overall user experience, enabling immersive virtual reality (VR) \cite{fantacci2021edge} and augmented reality (AR) applications \cite{chakrabarti2021deep}, as well as supporting advanced industrial applications such as remote surgery and autonomous manufacturing. As the world becomes increasingly connected, emerging wireless networks are paving the way for a truly interconnected and intelligent future \cite{shen2021holistic,sheemar2023full}.

Sophisticated antenna systems are an essential component for emerging 6G wireless systems, enabling network services across various frequencies and ranges\cite{sheemar2022practical,hajiyat2021antenna}. Over the years, antenna technologies have evolved significantly, driven by the growing demand for high-speed, high-bandwidth communication systems with improved performance, efficiency, and reliability \cite{ikram2022road,sheemar2021hybrid_PC,sheemar2022near,bellofiore2002smart,sheemar2021hybrid}. From traditional fixed antennas to adaptive and reconfigurable antennas \cite{sheemar2022hybrid,costantine2015reconfigurable,shlezinger2021dynamic}, the evolution of antenna technologies has led to significant advancements in wireless communication systems. With the emergence of new applications foreseen for 6G, there is a need for novel antennas that can meet the ever-increasing demands of these systems. This has led to the development of new antenna technologies, including metamaterials \cite{dong2012metamaterial}, and millimeter wave antennas \cite{sheemar2021hybrid}, among others.

To fully characterize the performance of the communication systems, it is first essential to investigate the radiation properties of the deployed antenna systems, i.e. perform antenna pattern characterization (APC), which captures information about several parameters, such as directivity, antenna gain and beamwidth, etc \cite{foegelle2002antenna}. Traditionally, a measurement campaign must be conducted by taking measurements at different angles in the range of $[0,2\pi]$ to achieve accurate APC. However, the characterization of such a crucial parameter is not an easy task due to significant challenging arising in the measurement campaign. Besides, for the evolving complex antenna technologies for 6G which are expected to create very beams to serve the users, naive measurements only based strategies result to be very time-consuming, and expensive, as they necessitate a dedicated anechoic chamber \cite{rodriguez2006open}  to nullify the effect of multi-path (MP) and electromagnetic interference, which can potentially lead to an accurate measurement campaign. Furthermore, 
in complex environments outside the anechoic chambers, the traditional APC techniques are prone to fail as they are incapable of distinguishing the total received information, i.e. from reflections or interference. This motivates the design of new APC methods which take into account the reflecting and/or interference in complex environments and intelligently process the measured data to extract information about the complex antenna patterns, which will be inherent to 6G communications.

In this paper, we propose to exploit signal-processing tools \cite{9414995,sheemar2021parallel} to refine the measurements taken in the MIMO system in a challenging environment, which leads to accurate APC. The antenna pattern (AP) dictates the effective power
irradiation in different directions. Consequently, in the direction where measurements are being conducted, the effective MIMO channel response results are affected by a scale factor, capturing the potential of a multi-antenna system in radiating different amounts of power in different directions. For an isotropic antenna, the scale factor results to be one, as the same amount of power is irradiated in each direction. To yield accurate APC for our antenna system, we first propose a minimum mean squared error (MMSE) estimator \cite{neumann2018learning} for the scaled channel response which captures the power irradiations efficiency of the multi-antenna system in the direction of measurements. Then the relationship between the effective scaled channel response and the line-of-sight (LoS) and non-LoS channel response is exploited to refine the measurements to jointly estimate the MP and the antenna pattern. The proposed approach is a joint and adaptive approach which must the executed for each position where the measurements are taken. Simulation results show that the proposed design achieves significant performance improvement in terms of APC accuracy. The performance improves significantly as the transmit power at the base station (BS) increases.

 The rest of the paper is organized as follows: We first present the system model and problem formulation in Section \ref{sistema}. The joint APC and MP characterization approach are proposed in Section \ref{soluzione}. Finally, Sections \ref{simulazioni} and \ref{conclusioni} present the simulation results and conclusions, respectively.



\section{System Model} \label{sistema}

\begin{figure}
    \centering
    \includegraphics[width=9cm,height=5cm]{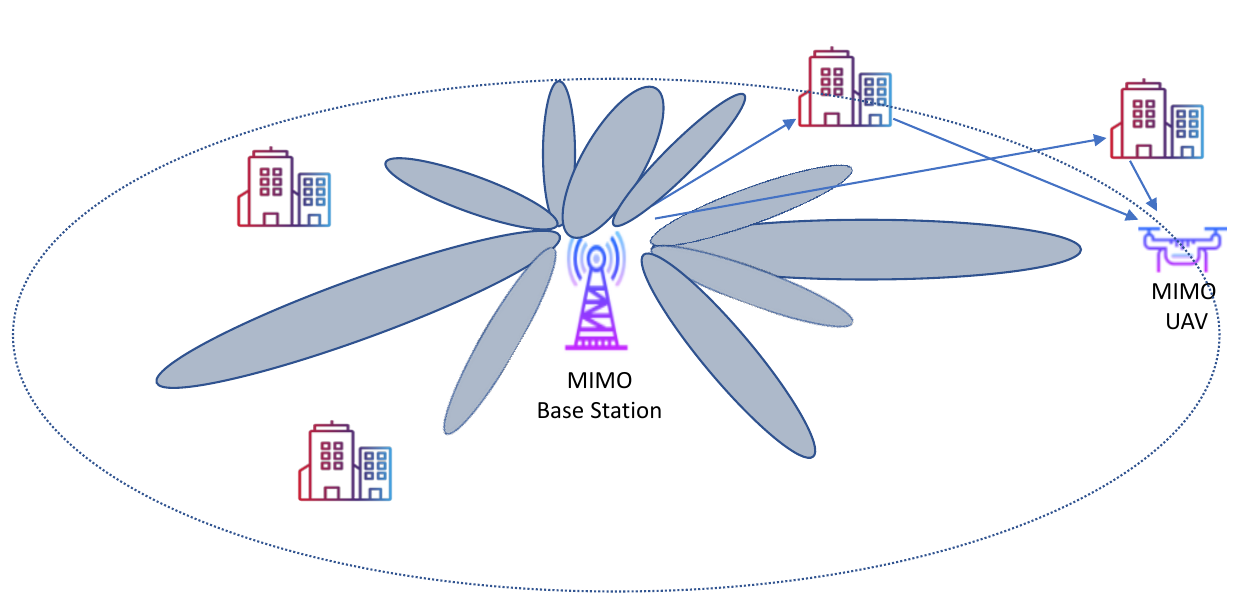}
    \caption{The measurement setup consisting of a MIMO BS and a flying UAV in the presence of reflections.}
    \label{UAV_scenario}
\end{figure}

In what follows, we consider the case of a one MIMO BS consisting of $N_{tx}$ transmit antennas, deployed at the height $h$ from the ground in an outdoor environment, as shown in Fig. \ref{UAV_scenario}. We assume that an Unmanned aerial vehicle (UAV) with $N_{rx}$ receive antennas, is flying at the same height $h$ with a circular trajectory of radius $d$, aiming at estimating the antenna pattern of the multi-antenna array deployed at the MIMO BS. The outdoor environment is assumed to contain reflectors which contribute to the MP. 
To simplify the analysis, we will disregard the takeoff and landing phases of the UAV. Additionally, we assume that the objective of the UAV is to complete a full circle of $360^\circ$, meaning that the initial position $\bmq_{s} = (x_{s},y_{s})^T$ and the destination position $\bmq_{d} = (x_{d},y_{d})^T$ at height $h$ are identical, i.e., $\bmq_{s} = \bmq_{d}$. The trajectory of the UAV at height $h$ is predetermined based on the surrounding environment of the BS. In the presence of obstacles, the radial distance $d$ can be adjusted to ensure a line-of-sight (LoS) path between the UAV and the BS is always maintained.

Let's consider the BS positioned at the center of a three-dimensional coordinate system, and denote $\theta_i$ as the angle between the BS and the UAV. The complete trajectory, which consists of a single circle, is divided into $\Theta$ evenly spaced points where the UAV pauses briefly to gather measurement data. Hence, we have $ \theta_1 \leq \theta_i \leq \theta_{\Theta}$, and we can denote the position of the UAV on the two-dimensional coordinate system as $\bmq(\theta_i) = (x(\theta_i),y(\theta_i))^T$.

Let us consider a flat block-fading MIMO system. \textcolor{black}{We assume the radiation pattern to be constant during the time for which the UAV takes the measurements.}
We adopt a pilot-based approach in which the UAV is aware of the signal being transmitted from the BS. Assume that for antenna pattern characterization, for each $\theta_i$ the BS transmits a sequence of $Q$ training samples with $Q > N_{t_x}$, collected in the matrix as $\bmP = (\bmp_1,......,\bmp_Q)$, with $\bmp_i \in \mathcal{C}^{N_{tx} \times 1} $. Moreover, $\bmP$ is supposed to be the same $\forall \;\theta_i$. In the case of an isotropic antenna system, the power radiated is equal in all directions when the BS transmits training samples $\bmP$. However, the evolving complex antenna systems for 6G and beyond will depend on beamforming techniques, enabling the allocation of a substantial power level in multiple desired directions using highly focused beams precisely aimed at the locations of the users.
Consequently, there can be a substantial discrepancy in the power radiated across various directions. Consider the scalar $a(\theta_1)$,
which represents the efficiency of power irradiation in the direction $\theta_i$.
The scalar $a$ satisfies the following conditions $0 \leq a_i(\theta_i) \leq 1$, where $a(\theta_i) =0$ and $a(\theta_i) =1$ denote the direction of radiation null or the main beam, respectively. Note that $a$ spanned over the interval $[0, 2 \pi]$ also represents the antenna pattern that we wish to estimate. Consequently, $\bmP(\theta_i)= (\bmp_1(\theta_i),......,\bmp_Q(\theta_i))$ denotes the effective power irradiated in the direction $\theta_i$, where $ \bmp_j(\theta_i)= \sqrt{a(\theta_i)} \bmp_j, \forall j$. While $\bmP$ is the same for each $\theta_i$, the effective irradiated power $\bmP(\theta_i)$ is different due to the non isotropic antenna array. Let $\bmY(\theta_i) \in \mathcal{C}^{N_{rx} \times Q}$ denote the effective received signal matrix at the UAV, which can be written as
\begin{equation} \label{signal_model}
    \bmY(\theta_i) = \bmH(\theta_i) \sqrt{\bma(\theta_i)} \bmP + \bmV(\theta_i),
\end{equation}
where $\bmV \in \mathcal{C}^{N_{rx} \times Q}$ denotes the noise.

\subsection{On the MIMO Channel Model for Antenna Pattern Characterization}
In a typical scenario, the MIMO 
channel $\bmH$ consists of line-of-sight (LoS) component and the MP components, denoted as $\bmH_{\text{LoS}}$ and $\bmH_{\text{MP}}$, respectively. 
For an isotropic antenna array, the LoS channel $\bmH_{\text{LoS}}$ is independent of $\theta_i$ as the path loss is the same for all points lying on the circular trajectory of the UAV of radius $d$. However, $\bmH_{\text{MP}}$ still depends on the angle $\theta_i$ as the total number of reflective paths can add up constructively or destructively depending on the position of the UAVs and reflectors. Hence, we can write

\begin{equation}
\bmH(\theta_i) = \bmH_{\text{LoS}}+ \bmH_{\text{MP}}(\theta_i)
\end{equation}
Given the aforementioned motivation, we can model the LoS channel as
 
\begin{equation} \label{scaling}
\bmH_{\text{LoS}}(\theta_i) = \sqrt{\alpha_{\text{LoS}}}\bma_r(\theta_r)\bma_t^H(\theta_i)
\end{equation}
where $\bma_r(\theta_r)$ and $\bma_t^H(\theta_i)$ denote the receive and transmit antenna array responses, respectively, and $\alpha_{\text{LoS}}$ denote the path-loss, which is the same for the UAV flying on the circular trajectory of radius $d$. The matrix  $\bmH_{\text{MP}}$ can be modelled as

\begin{equation}
    \bmH_{\text{MP}} = \sum_{k=1}^{L}\sqrt{\alpha_k(\theta_r)}\bma_r(\theta_i,k)\bma_t^H(\theta_i,k)
\end{equation}
where $L$ denote the total number of paths. 
By decomposing the effective channel in LoS and MP components, \eqref{signal_model} can be written as
\begin{equation}
\bmY(\theta_i) =  \sqrt{a(\theta_i)} \bmH_{\text{LoS}}(\theta_i) + \sqrt{a(\theta_i)} \bmH_{\text{MP}} \bmP +  \bmV(\theta_i)  .
\end{equation}
When the UAV performs measurements at the position $\theta_i$ for APC, the total measured power from the multi-antenna system is given by $||\bmY(\theta_i)||_F^2$, which is affected also by power absorbed from the multi-path. Ideally, to verify if the irradiation power satisfies the total power constraint $\gamma$, the MP contributions should be cancelled to measure the effective receive power irradiated due to direct irradiation, which results to be
\begin{equation} \label{problem}
p(\theta_i) = ||\bmY(\theta_i) - \sqrt{a(\theta_i)} 
 \bmH_{\textbf{MP}}(\theta_i)  \bmP||_F^2.
\end{equation}
 Adopting this approach is of extreme interest for higher frequencies such as millimeter wave, where the power from reflections becomes comparable to the LoS component.

\section{Problem Formulation and Solution} \label{soluzione}






To accurately estimate the antenna pattern, we propose to first estimate the scaled channel response in the direction $\theta_i$, denoted as $\bmH_a(\theta_i) = \sqrt{\bma(\theta_i)} \bmH(\theta_i)$ in the following. To estimate the scaled channel response, we aim at finding its minimum MSE estimator, for which the optimization problem can be formulated as

\begin{subequations}\label{WSR_problem}
\begin{equation}
\underset{\substack{\bmH}}{\min} \quad  ||\bmY(\theta_i) - \bmH_a(\theta_i)  \bmP + \bmV(\theta_i)||_F^2
\end{equation} 
\end{subequations}

By solving the problem above, we get the following optimal MMSE estimator for the scaled channel response

\begin{equation} \label{scaled_channel}
    \hat{\bmH}_a(\theta_i) = \bmY(\theta_i) \bmP^H (\bmP \bmP^H)^{-1}
\end{equation}

Given the scaled channel estimate which depends on the antenna array factor, the following equation holds
\begin{equation} \label{final_est}
    \hat{\bmH}_a(\theta_i) = \sqrt{\hat{a}(\theta_i)} \hat{\bmH}_{\text{LoS}} + \sqrt{\hat{a}(\theta_i)} \hat{\bmH}_{\text{MP}}(\theta_i)
\end{equation}
where $\hat{\bmH}_{\text{LoS}}$ and $\hat{\bmH}_{\text{MP}}$ denote the estimates for the LoS and multi-path component and $\sqrt{\hat{a}(\theta_i)}$ denotes the estimated components of the antenna array at position $\theta_i$. It is noteworthy that given the position of the BS with respect to UAV is always known which lies on a circle of radius $d$, $\hat{\bmH}_{\text{LoS}}$ can be easily obtained, which does not vary during the whole trajectory of the UAV. Given such information, we must find jointly $\hat{\bmH}_{\text{MP}}(\theta_i)$ and $a(\theta_i)$ given the scaled MMSE channel estimate. To do so, we consider minimizing the error between $\hat{\bmH}_a(\theta_i)$ and $\hat{\bmH}_{\text{MP}}(\theta_i)$ and $a(\theta_i)$, for which the MSSE optimization problem can be stated as

\begin{equation}  \label{opt_restated}
\underset{\substack{a(\theta_i),\bmH_{\text{MP}}(\theta_i)}}{\min} \quad   ||\hat{\bmH}_a(\theta_i)  - \sqrt{a(\theta_i)} \hat{\bmH}_{\text{LoS}} - \sqrt{a(\theta_i)} \bmH_{\textbf{MP}}(\theta_i) ||_F^2.
\end{equation}
We adopt an alternating optimization approach to iteratively optimize the values of $a(\theta_i),\bmH_{\text{MP}}(\theta_i)$. Note that the values of the scale factor satisfy $0 \leq a(\theta_i)\leq 1$. At the first position $\theta_1$ where the UAV takes the measurements, consider selecting the starting value of the scale factor $a(\theta_1)^{(0)} \in $ $[0,1]$. For the positions with $i\neq 1,$ the starting value of $a(\theta_i)^{(0)}$ can be chosen as $a(\theta_{i})^{(0)} = \hat{a}(\theta_{i-1})$, i.e. the one estimated at the previous position. Alternatively, the MP $\bmH_{\textbf{MP}}(\theta_i)$ can be first initialized. 

Given $a(\theta_1)^{(0)}$, we consider optimizing the estimate of $\bmH_{\textbf{MP}}(\theta_1)$ at the first iteration, denoted as $ \hat{\bmH}_{\textbf{MP}}(\theta_1)^{(1)}$. For such a purpose, we take the derivative of the objective function $\eqref{opt_restated}$ with respect to the conjugate of  $\bmH_{\textbf{MP}}(\theta_1)^{(1)}$, which leads to the following closed-form solution

 \begin{equation} \label{first_est}
     \hat{\bmH}_{\textbf{MP}}(\theta_1)^{(1)} = \frac{1}{\sqrt{a(\theta_1)^{(0)}}} \hat{\bmH}_a - \hat{\bmH}_{LoS}
 \end{equation}

Given the recently computed estimate for 
$\hat{\bmH}_{\textbf{MP}}(\theta_1)^{(1)}$, we aim at finding the optimal antenna pattern value in position $\theta_1$, by solving the following optimization problem

\begin{equation}  \label{opt_restated_scalar}
\underset{\substack{a(\theta_i)^{(1)}}}{\min} \quad   ||\hat{\bmH}_a(\theta_i)  - \sqrt{a(\theta_i)^{(1)}} \hat{\bmH}_{\text{LoS}} - \sqrt{a(\theta_i)^{(1)}} \hat{\bmH}_{\textbf{MP}}(\theta_1)^{(1)} ||_F^2.
\end{equation}

We consider solving this problem by performing a linear search for $a(\theta_i)^{(1)}$, restricted to the interval $[0,1]$, aiming at finding the scalar leading the MMSE. This leads to a simplified search-based approach which does not rely on heavy computations. Once the optimal $\hat{a}(\theta_i)^{(1)}$ has been found, the process can be repeated iteratively for both variables until convergence. Then the process must be repeated at each $\theta_i$ by first collecting the measurement data $\bmY(\theta_i)$ for which as the initial estimate for $a(\theta_i)^{(0)}$ the previously estimated value can be used. The overall procedure to find the optimal antenna pattern for the circular trajectory for the UAV is formally stated in Algorithm $\ref{alg_1}$.

\begin{algorithm}[t]  
\caption{Antenna Pattern Characterization}\label{alg_1}
\textbf{Initialize:} Set the training sequence $\bmP$, the measurement points  \quad $\theta_i$ and estimate $\hat{\bmH}_{\text{LoS}}$. \\
\textbf{for} $i=1:1:\Theta$
\begin{algorithmic}
\STATE \hspace{0.02cm} Get the measurements $\hat{\bmY}(\theta_i)$\\  
\STATE \hspace{0.02cm} Set $\hat{a}(\theta_i)^{(0)}$ \\
\STATE \hspace{0.02cm} Set $n=1$\\
\STATE \hspace{0.04cm} \textbf{Repeat until convergence} \\
\STATE \hspace{0.3cm} Estimate $ \hat{\bmH}_{\textbf{MP}}(\theta_1)^{(n)}$ with  $\eqref{first_est}$.
\STATE  \hspace{0.3cm} Find the optimal $\hat{a}(\theta_i)^{(n)}$ with linear search.
\STATE \hspace{0.3cm} n = n+1
\STATE  \hspace{0.001cm} \textbf{Save} in $a(\theta_i)$.\\
\STATE  \hspace{0.001cm} \textbf{Save} the effective power as \eqref{problem}, if needed.\\
\end{algorithmic} 
\textbf{end} \\
\label{Alg_1} \vspace{-3mm}
\end{algorithm}

 \section{Simulations Results} \label{simulazioni}

In this section, we present the simulation results to evaluate the performance of the proposed signal processing-based MIMO antenna pattern characterization technique.

\textcolor{black}{We consider an outdoor environment and we assume that the BS and the UAV deploy phased antenna arrays with the number of transmit and receive antennas $N_{tx}=10$ and $N_{rx}=8$, respectively.} A pilot sequence length $Q=100$ is assumed to be transmitted at each $\theta_i$. \textcolor{black}{We assume that the BS transmit at the rate of $R = 25\; symbols/sec$, which requires the UAV to collect measurements at each position for $4~$s. The UAV is assumed to take measurements on $\Theta = 50$ points equally distributed on the circular trajectory. By ignoring the flying time between two consecutive points to be negligible, the total flying time for the UAV results to be $\sim 200$s ($3.33~$mins).} The pilot sequence is designed with independent rows, which has been shown in the literature to perform optimally. We define the signal-to-noise-ratio (SNR) of our system as the transmit SNR, i.e., 
\begin{equation}
    \mbox{SNR}=\frac{\gamma}{\sigma^2},
\end{equation}
with $\gamma$ and $\sigma^2$ denoting the transmit power and the noise variance, respectively. We consider selecting $\gamma=1$ and selecting the noise variance to meet the transmit SNR requirement. The radius of the trajectory is $d=50~$m. We consider that the measurements take place for a BS deployed in sub $6$-GHz, for which the LoS is assumed to be dominant compared to the MP case. The MIMO channel response is modelled as a Rician fading channel model with Rician factor $\kappa =5~$dB.

\begin{figure}
  \centering
\includegraphics[width=0.49\textwidth,height=5.5cm]{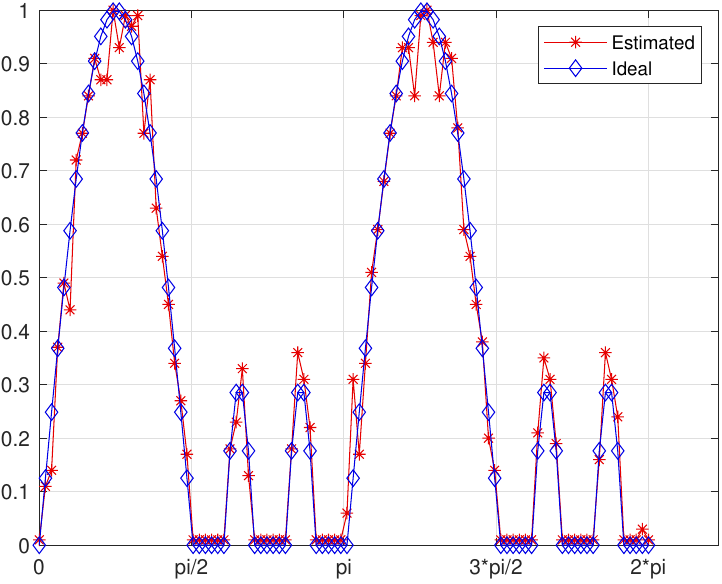}
     \caption{Estimated antenna pattern design at SNR=$10~$dB..}
     \label{fig:enter-label_1}
\end{figure}


In Figure \ref{fig:enter-label_1}, we present the performance analysis of the proposed approach for estimating the AP in a wireless communication system, specifically focusing on a SNR of $10~$dB. The obtained results clearly demonstrate the effectiveness and efficiency of our proposed approach in accurately estimating the AP. However, it is essential to note that there is a discernible mismatch observed at the estimated peaks of the AP curve, indicating a relatively poorer estimation in those instances. This mismatch can be attributed to the deliberate selection of a higher noise variance during the reporting of the results. The inclusion of these intentionally designed conditions enables a comprehensive evaluation of the performance of our proposed approach.

\begin{figure}
    \centering
\includegraphics[width=0.49\textwidth,height=5.5cm]{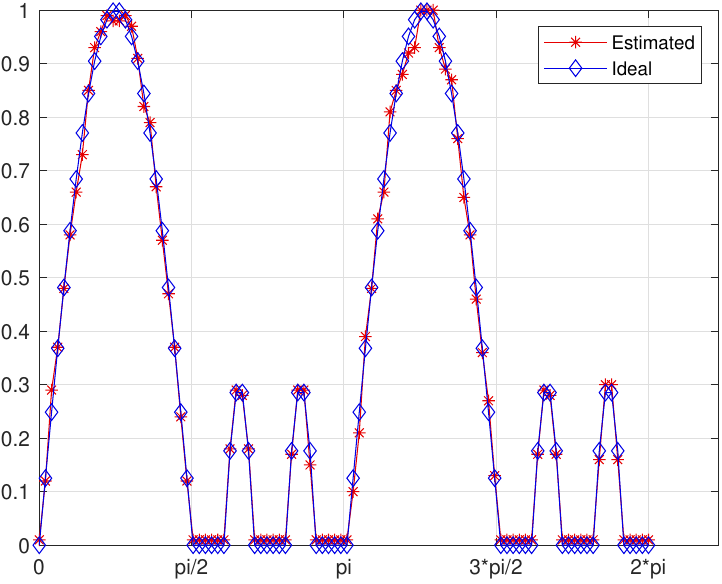}
     \caption{Estimated antenna pattern design at SNR=$20~$dB.}
     \label{fig:enter-label_2}
\end{figure}

Continuing our investigation, in Figure \ref{fig:enter-label_2}, we further explore the performance of the proposed scheme in estimating the AP, this time at a higher SNR of $20~$dB. The graphical representation clearly showcases the significant gains achieved by our proposed scheme in terms of accurately estimating the AP. Comparing this scenario to the previous case, we can observe a noteworthy reduction in the occurrence of erroneous estimates at the peaks, leading to a more precise estimation of the AP.

\begin{figure}
      \centering
\includegraphics[width=0.49\textwidth,height=5.5cm]{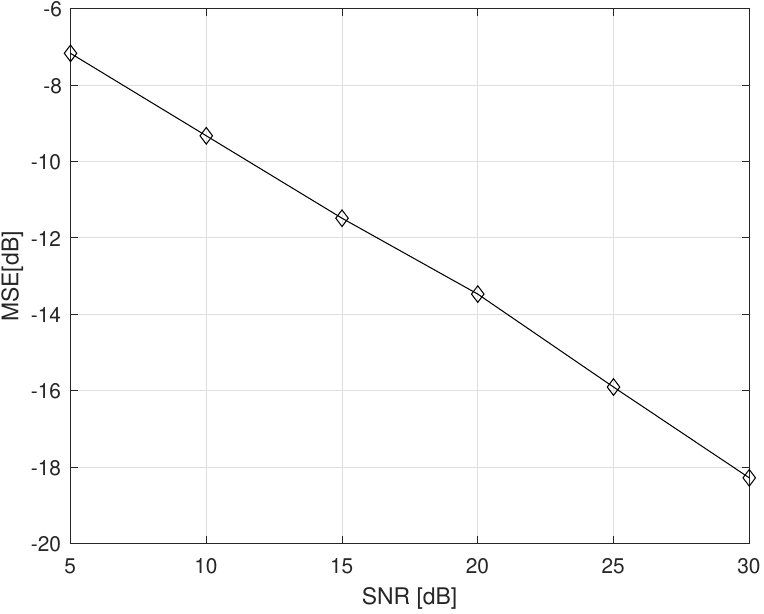}
     \caption{MSE as a function of the transmit SNR.}
     \label{mse}
 \end{figure}

To provide a quantitative analysis, we examine the MSE in dB between the estimated AP and the ideal AP and its behaviour as a function of the SNR. In Figure \ref{mse}, the plot depicts how the MSE decays as the SNR increases. It is evident that the MSE starts relatively high at low transmit SNR and rapidly decreases as the noise variance reduces. This indicates that our proposed approach exhibits improved accuracy in estimating the AP with reduced noise levels.


\begin{figure}
      \centering
\includegraphics[width=0.49\textwidth,height=5.5cm]{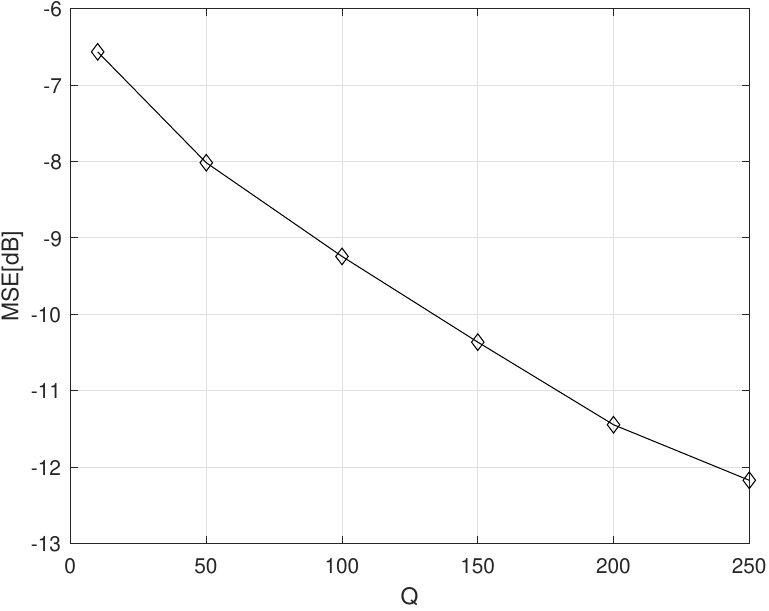}
     \caption{MSE as a function of the $Q$.}
     \label{mse_Q}
 \end{figure}

Furthermore, to explore potential enhancements in scenarios with limited transmit SNR, we consider the impact of increasing the length of the pilot sequence $Q$. Figure \ref{mse_Q} demonstrates the effect of varying pilot sequence lengths on reducing MSE at an SNR of $10~$dB. The results illustrate that a larger pilot sequence can compensate for the lower irradiating power, effectively reducing the MSE between the estimated and ideal AP.

In conclusion, our investigation highlights the effectiveness and efficiency of the proposed approach in accurately estimating AP in wireless communication systems. While some challenges arise in specific conditions with higher noise levels at the peaks of the AP, overall our algorithm still performs well. Overall, we can conclude that our proposed scheme exhibits significant gains in AP estimation accuracy at higher SNRs, and the MSE analysis confirms its improved performance as the noise variance reduces. Furthermore, we learned that the use of longer pilot sequences can further enhance the accuracy of AP estimation, particularly in scenarios with limited transmit SNR.

\section{conclusions} \label{conclusioni}
 In this paper, to achieve an accurate APC, we introduce a novel signal processing approach. By modelling the effect of antenna irradiation efficiency, we map the problem to a scaled MIMO channel estimation which captures the AP coefficient. Then, an MMSE estimator for the scaled channel response is proposed.  By utilizing this estimator, we establish a relationship between the effective scaled channel response and both the LoS and non-LoS channel responses. Building on this relationship, we refine the measurements and jointly estimate the MP and the antenna pattern. Our approach is characterized by its joint and adaptive nature, requiring execution for each position where measurements are taken. This adaptability allows us to account for variations in the environment and optimize the estimation process accordingly.
 Simulation results demonstrate the efficacy of our proposed design, showcasing significant improvements in terms of APC accuracy.

\bibliographystyle{IEEEtran}
\bibliography{FD_IRS}

\begin{thebibliography}{10}
\providecommand{\url}[1]{#1}
\csname url@samestyle\endcsname
\providecommand{\newblock}{\relax}
\providecommand{\bibinfo}[2]{#2}
\providecommand{\BIBentrySTDinterwordspacing}{\spaceskip=0pt\relax}
\providecommand{\BIBentryALTinterwordstretchfactor}{4}
\providecommand{\BIBentryALTinterwordspacing}{\spaceskip=\fontdimen2\font plus
\BIBentryALTinterwordstretchfactor\fontdimen3\font minus
  \fontdimen4\font\relax}
\providecommand{\BIBforeignlanguage}[2]{{%
\expandafter\ifx\csname l@#1\endcsname\relax
\typeout{** WARNING: IEEEtran.bst: No hyphenation pattern has been}%
\typeout{** loaded for the language `#1'. Using the pattern for}%
\typeout{** the default language instead.}%
\else
\language=\csname l@#1\endcsname
\fi
#2}}
\providecommand{\BIBdecl}{\relax}
\BIBdecl

\bibitem{sheemar2021game}
C.~K. Sheemar, L.~Badia, and S.~Tomasin, ``Game-theoretic mode scheduling for
  dynamic tdd in 5g systems,'' \emph{IEEE Communications Letters}, vol.~25,
  no.~7, pp. 2425--2429, 2021.

\bibitem{thomas2019multi}
C.~K. Thomas, C.~K. Sheemar, and D.~Slock, ``Multi-stage/hybrid bf under
  limited dynamic range for ofdm fd backhaul with mimo si nulling,'' in
  \emph{2019 16th International Symposium on Wireless Communication Systems
  (ISWCS)}.\hskip 1em plus 0.5em minus 0.4em\relax IEEE, 2019, pp. 96--101.

\bibitem{jiang2021road}
W.~Jiang, B.~Han, M.~A. Habibi, and H.~D. Schotten, ``The road towards {6G}: A
  comprehensive survey,'' \emph{IEEE Open Journal of the Communications
  Society}, vol.~2, pp. 334--366, 2021.

\bibitem{sheemar2020receiver}
C.~K. Sheemar and D.~Slock, ``Receiver design and agc optimization with self
  interference induced saturation,'' in \emph{IEEE ICASSP}, 2020, pp.
  5595--5599.

\bibitem{fantacci2021edge}
R.~Fantacci and B.~Picano, ``Edge-based virtual reality over {6G} terahertz
  channels,'' \emph{IEEE Network}, vol.~35, no.~5, pp. 28--33, 2021.

\bibitem{chakrabarti2021deep}
K.~Chakrabarti, ``Deep learning based offloading for mobile augmented reality
  application in {6G},'' \emph{Computers and Electrical Engineering}, vol.~95,
  p. 107381, 2021.

\bibitem{shen2021holistic}
X.~Shen, J.~Gao, W.~Wu, M.~Li, C.~Zhou, and W.~Zhuang, ``Holistic network
  virtualization and pervasive network intelligence for {6G},'' \emph{IEEE
  Communications Surveys \& Tutorials}, vol.~24, no.~1, pp. 1--30, 2021.

\bibitem{sheemar2023full}
C.~K. Sheemar, G.~C. Alexandropoulos, D.~Slock, J.~Querol, and S.~Chatzinotas,
  ``Full-duplex-enabled joint communications and sensing with reconfigurable
  intelligent surfaces,'' \emph{arXiv preprint arXiv:2306.10865}, 2023.

\bibitem{sheemar2022practical}
C.~K. Sheemar, C.~K. Thomas, and D.~Slock, ``Practical hybrid beamforming for
  millimeter wave massive {MIMO} full duplex with limited dynamic range,''
  \emph{IEEE OJ-COMS}, vol.~3, pp. 127--143, 2022.

\bibitem{hajiyat2021antenna}
Z.~R. Hajiyat, A.~Ismail, A.~Sali, and M.~N. Hamidon, ``Antenna in {6G}
  wireless communication system: Specifications, challenges, and research
  directions,'' \emph{Optik}, vol. 231, p. 166415, 2021.

\bibitem{ikram2022road}
M.~Ikram, K.~Sultan, M.~F. Lateef, and A.~S. Alqadami, ``A road towards {6G}
  communication—a review of {5G} antennas, arrays, and wearable devices,''
  \emph{Electronics}, vol.~11, no.~1, p. 169, 2022.

\bibitem{sheemar2021hybrid_PC}
C.~K. Sheemar and D.~Slock, ``Hybrid beamforming for bidirectional massive mimo
  full duplex under practical considerations,'' in \emph{IEEE VTC-Spring)},
  2021, pp. 1--5.

\bibitem{sheemar2022near}
C.~K. Sheemar, S.~Tomasin, D.~Slock, and S.~Chatzinotas, ``Near-field
  intelligent reflecting surfaces for millimeter wave mimo full duplex,''
  \emph{arXiv preprint arXiv:2211.10700}, 2022.

\bibitem{bellofiore2002smart}
S.~Bellofiore, C.~A. Balanis, J.~Foutz, and A.~S. Spanias, ``Smart-antenna
  systems for mobile communication networks. part 1. overview and antenna
  design,'' \emph{IEEE Antennas and Propagation Magazine}, vol.~44, no.~3, pp.
  145--154, 2002.

\bibitem{sheemar2021hybrid}
C.~K. Sheemar and D.~Slock, ``Hybrid beamforming and combining for millimeter
  wave full duplex massive {MIMO} interference channel,'' in \emph{IEEE
  GLOBECOM}, 2021, pp. 1--6.

\bibitem{sheemar2022hybrid}
C.~K. Sheemar, ``Hybrid beamforming techniques for massive mimo full duplex
  radio systems,'' Ph.D. dissertation, Ph. D. dissertation, EURECOM, 2022.

\bibitem{costantine2015reconfigurable}
J.~Costantine, Y.~Tawk, S.~E. Barbin, and C.~G. Christodoulou, ``Reconfigurable
  antennas: Design and applications,'' \emph{Proceedings of the IEEE}, vol.
  103, no.~3, pp. 424--437, 2015.

\bibitem{shlezinger2021dynamic}
N.~Shlezinger, G.~C. Alexandropoulos, M.~F. Imani, Y.~C. Eldar, and D.~R.
  Smith, ``Dynamic metasurface antennas for {6G} extreme massive {MIMO}
  communications,'' \emph{IEEE Wireless Communications}, vol.~28, no.~2, pp.
  106--113, 2021.

\bibitem{dong2012metamaterial}
Y.~Dong and T.~Itoh, ``Metamaterial-based antennas,'' \emph{Proceedings of the
  IEEE}, vol. 100, no.~7, pp. 2271--2285, 2012.

\bibitem{foegelle2002antenna}
M.~D. Foegelle, ``Antenna pattern measurement: concepts and techniques,''
  \emph{Compliance Engineering}, vol.~19, no.~3, pp. 22--33, 2002.

\bibitem{rodriguez2006open}
V.~Rodriguez, ``An open-boundary quad-ridged guide horn antenna for use as a
  source in antenna pattern measurement anechoic chambers,'' \emph{IEEE
  Antennas and Propagation Magazine}, vol.~48, no.~2, pp. 157--160, 2006.

\bibitem{9414995}
C.~K. Sheemar and D.~Slock, ``Beamforming for bidirectional {MIMO} full duplex
  under the joint sum power and per antenna power constraints,'' in \emph{IEEE
  ICASSP}, 2021, pp. 4800--4804.

\bibitem{sheemar2021parallel}
C.~K. Sheemar, S.~Chatzinotas, L.~E. Slock, Dirk, and J.~Querol, ``Parallel and
  distributed hybrid beamforming for multicell millimeter wave full duplex,''
  \emph{arXiv preprint arXiv:2112.02335}, 2021.

\bibitem{neumann2018learning}
D.~Neumann, T.~Wiese, and W.~Utschick, ``Learning the {MMSE} channel
  estimator,'' \emph{IEEE Transactions on Signal Processing}, vol.~66, no.~11,
  pp. 2905--2917, 2018.

\end{thebibliography}

\end{document}